\documentclass[prl,twocolumn,floatfix,nofootinbib,superscriptaddress]{revtex4}

\usepackage{amsmath}
\usepackage{amssymb}
\usepackage{amsthm}
\usepackage{bm}
\usepackage{epsfig}
\usepackage{graphicx}
\usepackage{longtable}
\usepackage{color}
\usepackage{epstopdf}
\usepackage{nicefrac}
\usepackage{tabularx}
\usepackage{dcolumn}  
\newcolumntype{L}[1]{>{\raggedright\arraybackslash}p{#1}}
\newcolumntype{C}[1]{>{\centering\arraybackslash}p{#1}}
\newcolumntype{R}[1]{>{\raggedleft\arraybackslash}p{#1}}

\usepackage[colorlinks=true, pdfstartview=FitV, linkcolor=purple, citecolor= purple,urlcolor=blue]{hyperref}
\definecolor{darkgreen}{rgb}{0,0.5,0}
\definecolor{purple}{rgb}{0.5,0,0.5}
\definecolor{nblue}{rgb}{0.0,0.0,0.50}
\definecolor{scarlet}{rgb}{1.0,0.2,0}

\begin{document}

\title{ $\bm{D^*}$ and $\bm{D^*_s}$ distribution amplitudes from Bethe-Salpeter wave functions}

\author{Fernando E. Serna}
\affiliation{LFTC, Universidade Cidade de S\~ao Paulo, Rua Galv\~ao Bueno 868, S\~ao Paulo, SP 01506-000, Brazil}
\affiliation{Departamento de F\'isica, Universidad de Sucre, Carrera 28 No. 5-267, Barrio Puerta Roja, Sincelejo, Colombia}

\author{Roberto Correa da Silveira}
\affiliation{LFTC, Universidade Cidade de S\~ao Paulo, Rua Galv\~ao Bueno 868, S\~ao Paulo, SP 01506-000, Brazil}

\author{Bruno El-Bennich}
\affiliation{LFTC, Universidade Cidade de S\~ao Paulo, Rua Galv\~ao Bueno 868, S\~ao Paulo, SP 01506-000, Brazil}


\begin{abstract}
We report on the first calculation of the longitudinal and transverse light front distribution amplitudes of the $D^*$ and $D^*_s$ mesons and their first four moments. 
As a byproduct, we also obtain these distribution amplitudes for the $\rho$, $\phi$, $K^*$ and $J/\Psi$ mesons and confirm a prediction of lattice QCD for the vector kaon:
while the longitudinal distribution amplitude is almost symmetric, the transverse one is oblique implying that the strange quark carries more momentum.
\end{abstract}

\date{\today}
\maketitle

\section{Motivation}

In relativistic quantum field theory the infinite degrees of freedom do not allow for a straightforward definition of a particle's wave function as in quantum mechanics. 
In particular, in Quantum Chromodynamics (QCD) the fundamental quark and gluon fields are not even observable. On the other hand, the bound states of 
\emph{valence\/} quark-antiquark pairs can be described by a Bethe-Salpeter wave function, the closest relative to a wave function in quantum mechanics. 
Still, in the instant-form of  QCD dynamics these wave functions are defined in an infinite-body field theory in which particles interact and their number is not conserved. 
 
One could overcome this difficulty if the hadron's light-front wave function was known exactly, though realistic calculations of hadronic bounds states in the front form 
are a challenging task~\cite{dePaula:2020qna}. A different path to a sensible definition of a wave function in quantum field theory is drawn by projecting the Bethe-Salpeter 
wave functions  in the instant form on the light front. Depending on the projection chosen this yields the hadron's light-front wave function or its light-front distribution amplitude 
(LFDA). The latter describes the longitudinal momentum distribution of valence quarks in the limit of negligible transverse momentum. While they are non-measurable 
objects, they are widely being applied in hadron and flavor physics. 

For instance, the asymptotic LFDA of the pion, $\phi(x,\mu)  \stackrel{\mu \to \infty }{=}   6x(1-x)$, enters in  the expression of its elastic electromagnetic form factor at 
very large momentum transfers~\cite{Efremov:1978rn,Lepage:1980fj}. Since the LFDAs are scale-dependent and become broader at smaller momenta, they directly 
influence the momentum dependence of the elastic form factors in momentum regions accessible in collider experiments~\cite{Chang:2013nia,Raya:2015gva,Raya:2019dnh}. 
Weak $B$ decays into two light(er) mesons are frequently treated as hard exclusive processes in which the decay amplitude is factorized into perturbative short-distance 
contributions and a nonperturbative transition amplitude. Here too, the LFDAs enter both, the hard-scattering integrals and the heavy-to-light transition 
amplitudes~\cite{Beneke:2002jn,Bauer:2005wb,El-Bennich:2006rcn,El-Bennich:2009gqk,Leitner:2010fq}. More recently, the exclusive electroweak production of 
$D_s^{(*)}$ mesons on an unpolarized nucleon was investigated in the framework of collinear QCD factorization which also involves the heavy meson's 
LFDA~\cite{Pire:2015iza,Pire:2017lfj,Pire:2017yge,Pire:2021dad}.

Beyond its numerous applications in hard exclusive processes, these one-dimensional distributions provide a practical probability interpretation of partons, as in this frame 
the particle number is conserved. Namely, the distributions $\phi(x,\mu)$ express the light-front fraction of the hadron's momentum that a valence quark carries. Another 
compelling feature is that one can observe the qualitative and quantitative impact of dynamical chiral symmetry breaking (DCSB) on the LFDA at a given scale $\mu$. 
For instance, the distribution amplitude $\phi_\pi (x,\mu)$ of the pion is a concave function which clearly evolves from its asymptotic $\mu \to \infty$ form to a much broader 
distribution~\cite{Chang:2013pq}. Similarly, the kaon's distribution amplitude, $\phi_K (x,\mu)$, is not symmetric about the midpoint $x = 1/2$, which expresses nothing 
but SU(3) flavor symmetry breaking, and that asymmetry is exacerbated with increasing mass difference of the quarks~\cite{Shi:2015esa,Serna:2020txe}. 

The question arises of how DCSB impacts antiquark-quark states in other $J^{PC}$ channels and an extension to the vector mesons is natural. 
Moreover, the LFDA of vector mesons arises in the collinear factorization of weak $B$-decay amplitudes~\cite{Beneke:2000wa} and in diffractive vector-meson 
production~\cite{Forshaw:2010py,Forshaw:2012im}. Within the combined framework of the Dyson-Schwinger equation (DSE) and the Bethe-Salpeter equation 
(BSE)~\cite{Bashir:2012fs} the LFDAs of the $\rho$ and $\phi$ mesons were calculated in Ref.~\cite{Gao:2014bca} and later the LFDAs of heavy quarkonia were obtained 
in Ref.~\cite{Ding:2015rkn}. In here, using a kindred DSE and BSE framework, we extend earlier work on $D$ and $D_s$ distribution amplitudes~\cite{Serna:2020txe} 
to those of their vector partners and make predictions for the twist-2 LFDA of the  $D^*$ and $D_s^*$ mesons considering the two-quark Fock-state of their light front 
wave function. Along the way, we compute the LFDA of the $\rho$, $K^*$ $\phi$ and $J/\Psi$ mesons and compare them with the distribution amplitudes of other 
approaches~\cite{Forshaw:2010py,Forshaw:2012im,Ball:2007zt,Boyle:2008nj,Hua:2020gnw,Ding:2015rkn}.
\vspace*{-3mm}


\section{Twist-Two Distribution Amplitudes}
\vspace*{-2mm}

A vector meson with total momentum $P$ and mass $m_V$, $P^2=-m_V^2$, made of a quark and an antiquark of flavors $f$ and $g$ is described by four twist-two 
distribution amplitudes, though only two of them are independent at leading twist as a consequence of a Wandzura-Wilczek type of relation~\cite{Ball:2007zt}. 
The two LFDAs we consider, $\phi^{\parallel}_V(x,\mu)$ and $\phi^{\perp}_V(x,\mu)$, describe the fraction of total momentum on the light front, $x= k^+/P^+ = 
(k_0+k_z)/(P_0+P_z)$, carried  by the quark in longitudinally and transversely polarized mesons, respectively. They can be extracted from the Bethe-Salpeter wave 
function, $\chi^{fg}_{V\nu}(k,P)$,  with the following projections onto the light front~\cite{Gao:2014bca,Ding:2015rkn}:
\begin{align}
  f_V  \phi^\parallel_V(x,\mu)  =  \frac{m_VN_c{\cal Z}_2}{\sqrt{2}\  n\cdot P} &  \
       \mathrm{Tr}_D\!   \int^\Lambda  \!\! \frac{d^4k}{(2\pi)^4}  \,  \delta(n\cdot k_\eta-x\,n\cdot P)   \nonumber  \\   
       & \times \,  \gamma\cdotp n\,  n_\nu   \, \chi^{fg}_{V\nu}(k,P) \ ,    
\label{Par-LCDA}    
\end{align}
\begin{align}
    f^{\perp}_V  \phi^\perp_V(x,\mu)  =  -\frac{N_c{\cal Z}_T}{2\sqrt{2}} & \
     \mathrm{Tr}_D\!  \int^\Lambda  \! \!\frac{d^4k}{(2\pi)^4} \, \delta(n\cdot k_\eta-x\,n\cdot P)  \nonumber    \\   
     &  \times  \, n_\mu\sigma_{\mu\rho} \, \mathcal{O}^{\perp}_{\rho\nu}\, \chi^{fg}_{V\nu}(k, P) \ ,
 \label{Perp-LCDA}
\end{align}
where $N_c =3$, $n= (0,0,1,i)$ is a light-like vector and $\bar n =\tfrac{1}{2}(0,0,-1,i)$ its conjugate with $n^2 = \bar n^2=0$, $n\cdot P =-m_V$, 
$\bar n\cdot P=-m_V/2$
and $n\cdot \bar n = -1$.\footnote{We use Euclidean metric with the Dirac algebra:  $\{\gamma_\mu,\gamma_\nu\} = 2\delta_{\mu\nu}$; 
$\gamma_\mu^\dagger = \gamma_\mu$; $\gamma_5= \gamma_4\gamma_1\gamma_2\gamma_3$, tr$[\gamma_4\gamma_\mu\gamma_\nu\gamma_\rho\gamma_\sigma]=-4\, 
\epsilon_{\mu\nu\rho\sigma}$;  $\sigma_{\mu\nu}=(i/2)[\gamma_\mu,\gamma_\nu]$;  $a\cdot b = \sum_{i=1}^4 a_i b_i$. A time-like vector $P_\mu$ satisfies $P^2<0$. }  
In Eq.~\eqref{Perp-LCDA} the Dirac commutator $\sigma_{\mu\nu}$ is contracted with the tensor~\cite{Lu:2021sgg},
\begin{equation}          
          \mathcal{O}^{\perp}_{\rho\nu} = \delta_{\rho\nu}+n_\rho\bar n_\nu + \bar n_\rho n_\nu \ .
\end{equation} 
In Eqs.~\eqref{Par-LCDA} and \eqref{Perp-LCDA}, $\chi^{fg}_{V\nu}(k,P) = S_f(k_\eta)$ $\Gamma^{fg}_{V\nu}(k, P) S_g(k_{\bar\eta})$ is the projected wave function,
where $\Gamma^{fg}_{V\nu}(k,P)$ denotes the Bethe-Salpeter amplitude (BSA) and $S_f(k_\eta )$ and $S_g(k_{\bar \eta})$ are respectively the quark and antiquark 
propagators with momenta $k_{\eta}=k+\eta P$ and $k_{\bar\eta}=k-\bar\eta P$. The details of their calculation, solving numerically the DSE for the quarks of a given 
flavor and the BSE for a vector meson, in particular the $D$ and $D^*$ mesons, are provided elsewhere~\cite{Serna:2020txe,Mojica:2017tvh,El-Bennich:2021ldv}. 
The parameters  $\eta +\bar \eta =1$ define momentum fractions and $\Lambda$ is an ultraviolet regularization mass-scale; no observables can depend  on $\eta$, 
$\bar \eta$ and $\Lambda$ owing to Poincar\'e covariance. Furthermore, ${\cal Z}_2(\mu,\Lambda)$ is the wave-function renormalization constant and  
${\cal Z}_T(\mu,\Lambda)$ is the tensor-vertex renormalisation constant of the quark. Both constants as well as $f^{\perp}_V$ depend on the renormalization 
scale $\mu$, whereas $ f_V$ is renormalization-point  independent and measures the strength of the  $\rho^0 \to e^+ e^-$ decay amplitude. 

The expressions for $\phi^\parallel_V(x,\mu)$ and $\phi^\perp_V(x,\mu)$ in Eqs.~\eqref{Par-LCDA} and \eqref{Perp-LCDA} are not amenable to straightforward numerical 
integration. Instead, one computes Mellin moments~\cite{Chang:2013pq},
\begin{eqnarray}
   \langle x^m\rangle_{\parallel}   & = \int^1_0  x^m\,  \phi^{\parallel}_V(x,\mu) \, dx\ ,  
   \label{xmpara} \\
   \langle x^m\rangle_{\perp}       &  = \int^1_0 x^m\,  \phi^{\perp}_V(x,\mu) \,  dx \ , 
  \label{xmperp}
\end{eqnarray}
from which one can reconstruct the distribution amplitudes on the domain $x \in [0,1]$. The BSA normalization ensures that $ \langle x^0 \rangle_{\parallel} =
\langle x^0\rangle_{\perp} = 1$ which in turn defines the vector and tensor decay constants, $f_V^\parallel$ and $f^{\perp}_V$.

Integrating both sides of Eqs.~\eqref{Par-LCDA} and \eqref{Perp-LCDA} and applying the Dirac-function property $\int_{0}^{1}  x^{m} \delta(a-x b) dx = 
\frac{a^{m}}{b^{m+1}} \,\theta(b-a) $, leads to the expressions, 
\begingroup
\addtolength{\jot}{1em}
\begin{align}
   \langle x^m\rangle_{\parallel}   =  \frac{m_VN_c{\cal Z}_2}{\sqrt{2}\, f_V} & \, \mathrm{Tr}_D  \! \int^\Lambda \! \frac{d^4k}{(2\pi)^4}
                     \frac{(n\cdot k_\eta)^m}{(n\cdot P)^{m+2}}   \nonumber \\ 
                     & \times  \, \gamma\cdot n\,n_\nu  \, \chi^{fg}_{V\nu}(k,P)\, ,   
\label{Mom-Par-LCDA}  \\  
   \langle x^m\rangle_{\perp} =   -\frac{N_c{\cal Z}_T}{2\sqrt{2}\, f^{\perp}_V} & \, \mathrm{Tr}_D \! \int^\Lambda \! \frac{d^4k}{(2\pi)^4}
        \frac{(n\cdot k_\eta)^m}{(n\cdot P)^{m+1}} \nonumber \\
        & \times \, n_\mu\sigma_{\mu\rho}\, \mathcal{O}^{\perp}_{\rho\nu}  \, \chi^{fg}_{V\nu}(k,P) \, .  \hspace*{3mm}
\label{Mom-Perp-LCDA}
\end{align}
\endgroup
With this, we are in principle able to compute Mellin moments to arbitrary order $m$. We do so by employing the numerical solutions of the quark propagators for 
complex momenta defined by the parabolas, $k_\eta^2 =  k^2  - \eta^2 m^2_V  + 2 i \eta \, m_V  | k | z_k$ and $k_{\bar \eta}^2  =  k^2  - {\bar\eta}^2 m^2_V  
- 2 i\bar\eta \, m_V | k | z_k$, where $z_k = k\cdot P /|k||P|$, $-1 \leq z \leq +1$, and of the BSA of the vector mesons~\cite{El-Bennich:2021ldv}. That is, other 
than in Ref.~\cite{Serna:2020txe}, we do not rely on complex-conjugate pole parametrizations of the propagators nor on Nakanishi representations of the BSA,
as the latter introduce ambiguities when fitted to numerical solutions. However, direct integration comes at the price that we can only access moments up to 
$m_\mathrm{max} = 4-6$, as the numerical error of the integral becomes significant for larger moments. These moments, though, are sufficient to reconstruct 
the desired LFDA.

\begin{table*}[t!]
\renewcommand{\arraystretch}{1.2}
\setlength{\tabcolsep}{11pt}
\centering
\begin{tabular}{c|c|c|c|c||c|r|c}
\hline 
   &$\langle x\rangle_{\parallel,\perp}$ & $\langle x^2\rangle_{\parallel,\perp}$  &  $\langle x^3\rangle_{\parallel,\perp}$ & $\langle x^4\rangle_{\parallel,\perp}$ & 
   $a_1^{\parallel,\perp}$ & \multicolumn{1}{c}{$a_2^{\parallel,\perp}$} & $\alpha^{\parallel,\perp}$  \\ 
\hline 
   $\rho_\parallel$&0.500& 0.312& 0.226 & 0.161 & 0.0 & $0.003\pm0.038$&$0.908\pm0.023$   \\
   $\rho_\perp$& 0.500& 0.312& 0.218& 0.160  & 0.0 & $-0.136\pm0.007$&$0.799\pm0.006$ \\  
\hline
   $\phi_\parallel$&0.500&0.296& 0.195 & 0.134 & 0.0& $-0.372\pm0.010$&$0.864\pm0.010$  \\
   $\phi_\perp$& 0.500& 0.296& 0.193 & 0.134 & 0.0 &$-0.386\pm0.002$&$0.870\pm0.002$  \\
\hline
   $K^*_\parallel$&0.509&0.323& 0.236 & 0.179 & $0.041\pm0.027$& $-0.191\pm0.048$&$0.643\pm0.031$  \\
   $K^*_\perp$& 0.528& 0.351& 0.262& 0.204 &  $0.119\pm 0.003$& $0.122\pm0.015$&$0.840\pm0.019$  \\
\hline
\end{tabular}
\caption{The first four Mellin moments, $\langle x^m\rangle_{\parallel}$ and $\langle x^m\rangle_{\perp}$, of the light vector mesons and the coefficients of 
              their reconstructed  Gegenbauer expansion~\eqref{Gegen-form}. The errors on $a_1$, $a_2$ and $\alpha$ stem from the minimization. }
\label{x-moments-rho-phi-K}                
\end{table*}


\begin{table*}[t!]
\centering
\renewcommand{\arraystretch}{1.2}
\setlength{\tabcolsep}{10pt}
\begin{tabular}{l|c|c|c|c|c|c|c|c}
\hline
    $\rho_{\parallel,\perp}$  & $\langle \xi^2\rangle_\parallel$  & $\langle \xi^2\rangle_\perp$ &  $\langle \xi^4\rangle_\parallel$  &  $\langle \xi^4\rangle_\perp$ &
    $\langle \xi^6\rangle_\parallel$  & $\langle \xi^6\rangle_\perp$  & $\langle \xi^8\rangle_\parallel$  & $\langle \xi^8\rangle_\perp$ \\ 
\hline 
 Herein  &0.263 & 0.250& 0.136  & 0.127& 0.090  & 0.081 & 0.062  & 0.044  \\
 DSE~\cite{Gao:2014bca} & 0.231 & 0.252& 0.109 & 0.126& 0.065  & 0.079& 0.044 & 0.056  \\
 QCDSR~\cite{Ball:2007zt} & 0.234  & 0.238& 0.109  &  0.111 & 0.063  & 0.065  & 0.042  & 0.043  \\
 HERA~\cite{Forshaw:2010py,Forshaw:2012im} & 0.227 & 0.260 & 0.105  & 0.130 & 0.062  & 0.079 & 0.041  & 0.054  \\
 LQCD~\cite{Boyle:2008nj}  & 0.240(40)  &  &  & & &  \\

\hline\hline
   $\phi_{\parallel,\perp}$  &  $\langle \xi^2\rangle_\parallel$  & $\langle \xi^2\rangle_\perp$ &  $\langle \xi^4\rangle_\parallel$  &  $\langle \xi^4\rangle_\perp$ &
    $\langle \xi^6\rangle_\parallel$  & $\langle \xi^6\rangle_\perp$  & $\langle \xi^8\rangle_\parallel$  & $\langle \xi^8\rangle_\perp$ \\
\hline 
   Herein & 0.186  & 0.182  & 0.077  &  0.073&0.042  &  0.039  & 0.026  & 0.024  \\
   DSE~\cite{Gao:2014bca}   & 0.233 & 0.253 & 0.111 & 0.127 & 0.067 & 0.080 & 0.046 & 0.056 \\
   QCDSR~\cite{Ball:2007zt}  & 0.245 & 0.238 & 0.115 & 0.111& 0.068 & 0.065 & 0.045 & 0.043  \\
   LQCD~\cite{Hua:2020gnw}  & 0.212 & 0.250 & 0.097 & 0.127 & 0.057 & 0.081 & 0.039 & 0.058 \\
\hline \hline
$K^*_{\parallel,\perp}$  & $\langle \xi^2\rangle_\parallel$  & $\langle \xi^2\rangle_\perp$ &  $\langle \xi^4\rangle_\parallel$  &  $\langle \xi^4\rangle_\perp$ &
    $\langle \xi^6\rangle_\parallel$  & $\langle \xi^6\rangle_\perp$  & $\langle \xi^8\rangle_\parallel$  & $\langle \xi^8\rangle_\perp$ \\ 
\hline 
  Herein  &0.272 & 0.298 & 0.146  & 0.164 & 0.097  & 0.109 & 0.072  & 0.080  \\
  QCDSR~\cite{Ball:2007zt}  & 0.227  & 0.227 & 0.104 & 0.104  & 0.060 & 0.060 & 0.039 & 0.039 \\
  LQCD~\cite{Hua:2020gnw} & 0.200  & 0.292&0.088  & 0.162 & 0.050 & 0.111 & 0.032 & 0.084 \\
\hline 
\end{tabular}
\caption{Comparison of $\langle  \xi^{2m} \rangle_{\parallel,\perp}$ moments for the $\rho$, $\phi$ and $K^*$ mesons. The QCDSR values are obtained with 
              Eqs.~\eqref{xmpara} and \eqref{xmperp} employing the Gegenbauer expansion~\eqref{Gegen-form} with $\alpha =3/2$ and the value for $a_2$ in 
              Ref.~\cite{Ball:2007zt}. Similarly, we fit the tabulated values of $\phi^\parallel_V(x,\mu)$ and $\phi^\perp_V(x,\mu)$ provided in Ref.~\cite{Hua:2020gnw} 
              with the same Gegenbauer expansion and use them to calculate the moments.}
\label{xi-moments-rho-phi-K}  
\end{table*}


We proceed as in Refs.~\cite{Chang:2013pq,Shi:2015esa,Serna:2020txe,Gao:2014bca,Ding:2015rkn} and in the case of the light vector mesons
we use an expansion in terms of Gegenbauer moments $C_n^\alpha (2x-1)$, which form a complete orthonormal set on $x\in [0,1]$ with respect to the measure 
$[x(1-x)]^{\alpha-1/2}$, in order to reconstruct their two independent twist-two LFDAs ($\bar x = 1-x$):
\begin{equation}
   \phi^{\parallel,\perp}_{V \mathrm{rec.}} (x,\mu)  = \mathcal{N}(\alpha)[x\bar x]^{\alpha-\tfrac{1}{2}} \Bigg [1+\sum^N_{n=1}a_n C^\alpha_n(2x-1) \Bigg ]  .
\label{Gegen-form}
\end{equation}
This expansion is employed for neutral mesons as well as for flavored mesons, which are not $C$-parity eigenstates. In case of the former, the odd components $a_n$ 
vanish. In fitting the calculated moments in Eqs.~\eqref{Mom-Par-LCDA} and \eqref{Mom-Perp-LCDA}, we consider, besides the coefficients $a_n$, the power $\alpha$ 
itself a parameter rather than projecting on the $\alpha =3/2$ basis. This allows to limit the expansion to $N=2$ and considerably simplifies the fits discussed 
below~\cite{Chang:2013pq}. The normalization is obtained as,
\begin{equation}
    \mathcal{N}(\alpha)  = \frac{\Gamma(2\alpha+1)}{[\Gamma(\alpha+1/2)]^2} \ . 
\end{equation}
The heavy vector mesons, i.e. the $D^*$, $D^*_s$ and $J/\psi$, are parametrized with a different expression:
\begin{equation}
   \phi_{V\mathrm{rec.}}^{\parallel,\perp}(x,\mu) = \mathcal{N}(\alpha,\beta) \,  4x\bar x\,e^{4\,\alpha x\bar x+ \beta (x-\bar x)} \ .
  \label{phi-heavy} 
\end{equation}
This functional form is more appropriate for a distribution amplitude with a convex-concave-convex functional behavior that tends to a $\delta$-function in the infinite
heavy quark limit, as the use of an expansion, such as in Eq.~\eqref{Gegen-form}, leaves one no choice but to retain a large number of Gegenbauer moments. 
A very similar functional expression is also found when the Nakanishi weight function is extracted from the quarkonia's Bethe-Salpeter wave function~\cite{Gao:2016jka}. 
The normalization is given by~\cite{Serna:2021xnr}, 
\begin{align}
   \mathcal{N}(\alpha, \beta ) &  =  \ 16\,\alpha^{5/2} \Bigg [ 4\sqrt{\alpha}\, \left ( \beta \sinh \beta +2\alpha \cosh \beta \right )  \nonumber \\ 
           + &  \ \sqrt{\pi}\,  e^{\alpha+\frac{\beta^2}{4\alpha}} \left ( -2\alpha+4\alpha^2 - \beta^2 \right )  \nonumber \\ 
   \times & \,  \Bigg \{ \operatorname{Erf}  \left(\frac{2\alpha-\beta}{2\sqrt{\alpha}}\right) + \operatorname{Erf} \left( \frac{2\alpha+\beta}{2\sqrt{\alpha}} \right ) \Bigg \}  \Bigg ]^{-1} 
   ,
\end{align}
in which the error function is defined as: $\operatorname{Erf}(x)  = \frac{2}{\sqrt{\pi}}\int^x_0 dt\, e^{t^2}$. 

We thus reconstruct the vector LFDAs by minimizing the sum, 
\begin{equation}
   \epsilon_{\parallel,\perp} = \sum^{m_\textrm{max}}_{m=1}  \left | \frac{\langle x^m \rangle^\textrm{rec.}_{\parallel, \perp}}{\langle x^m\rangle_{ \parallel,\perp}} -1 \right | \ ,
\label{reco}
\end{equation}
where the moments $\langle x^m \rangle^\textrm{rec.}_{\parallel, \perp}$ are calculated using Eqs.~\eqref{xmpara} and \eqref{xmperp} and the expansion in either 
Eq.~\eqref{Gegen-form} or Eq.~\eqref{phi-heavy}, whereas $\langle x^m\rangle_{ \parallel,\perp}$ denotes the moments in Eqs.~\eqref{Mom-Par-LCDA} and 
\eqref{Mom-Perp-LCDA}. It is useful to contrast our predictions for the longitudinal and transverse LFDAs with those obtained using other approaches, namely 
with lattice QCD (LQCD)~\cite{Boyle:2008nj,Hua:2020gnw}, QCD sum rules (QCDSR)~\cite{Ball:2007zt} and with earlier calculations in the DSE-BSE 
framework (DSE)~\cite{Gao:2014bca}. In order to do so we also compute the moments,
\begin{equation}
\label{moments-v2}
 \left  \langle  \xi^{2m} \right  \rangle_{\parallel,\perp}  =  \int_{0}^{1}  \, \xi^{2m} \, \phi^{\parallel,\perp} _V (x, \mu ) dx \ ,
\end{equation}
in terms of the difference of momentum fractions, $\xi = x-(1-x) = 2 x-1$.


%
\begin{figure}[t!] 
\centering
\vspace*{-6mm}
  \includegraphics[scale=0.45,angle=0]{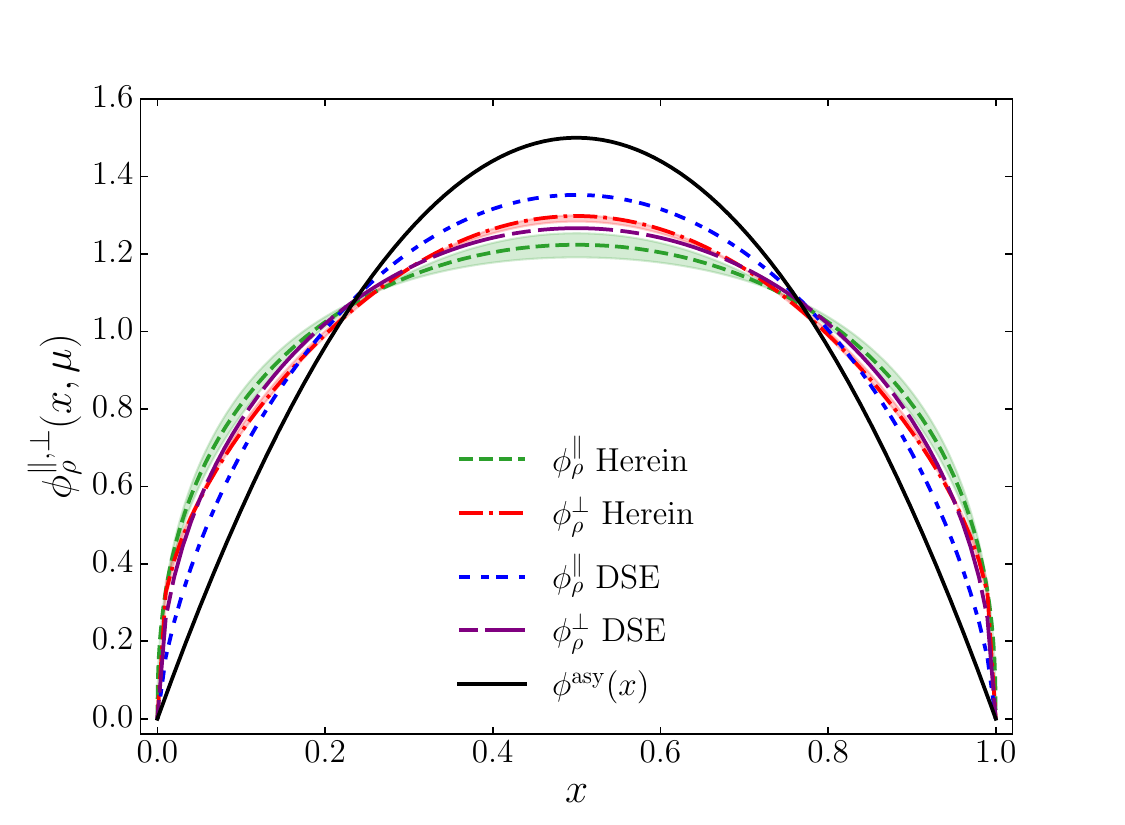} 
  \includegraphics[scale=0.45,angle=0]{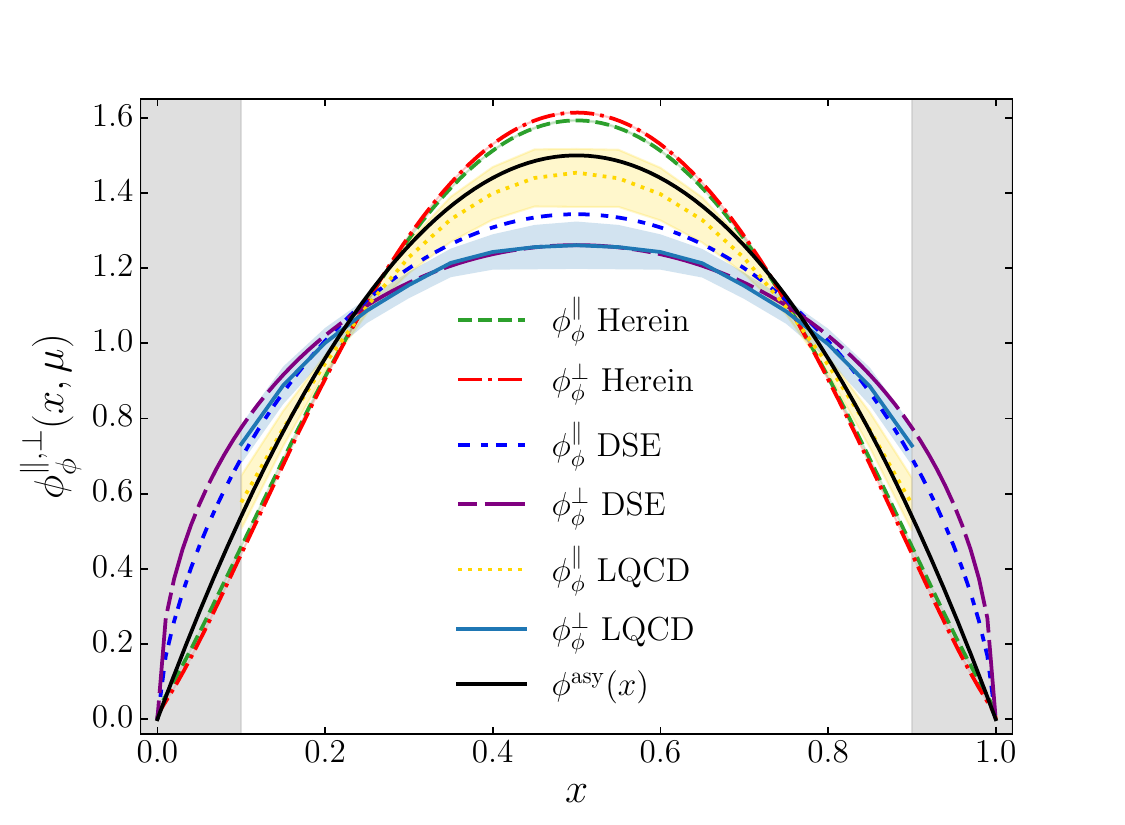}  
 \caption{Comparison of the longitudinal and transverse LFDAs for the $\rho$ ({\em top panel\/}) and $\phi$ ({\em bottom panel\/}) mesons with those of 
               Refs.~\cite{Gao:2014bca} (DSE) and \cite{Hua:2020gnw} (LQCD) at $\mu = 2$~GeV. Error bands reflect the uncertainties of the fit parameters in 
               Table~\ref{x-moments-rho-phi-K}. The intervals $0 \leq x < 0.1$, $1 \geq x > 0.9$ are shaded, as LQCD does not provide data for these momentum 
               fractions due to systematic errors. For comparison, we plot the asymptotic LFDA $\phi(x,\mu)  \stackrel{\mu \to \infty }{=}   6x \bar x$. }
 \label{fig:Compa-LCDAs-rho-phi} 
\end{figure}
%
%
\begin{figure}[t] 
\centering
\vspace*{-6mm}
  \includegraphics[scale=0.433,angle=0]{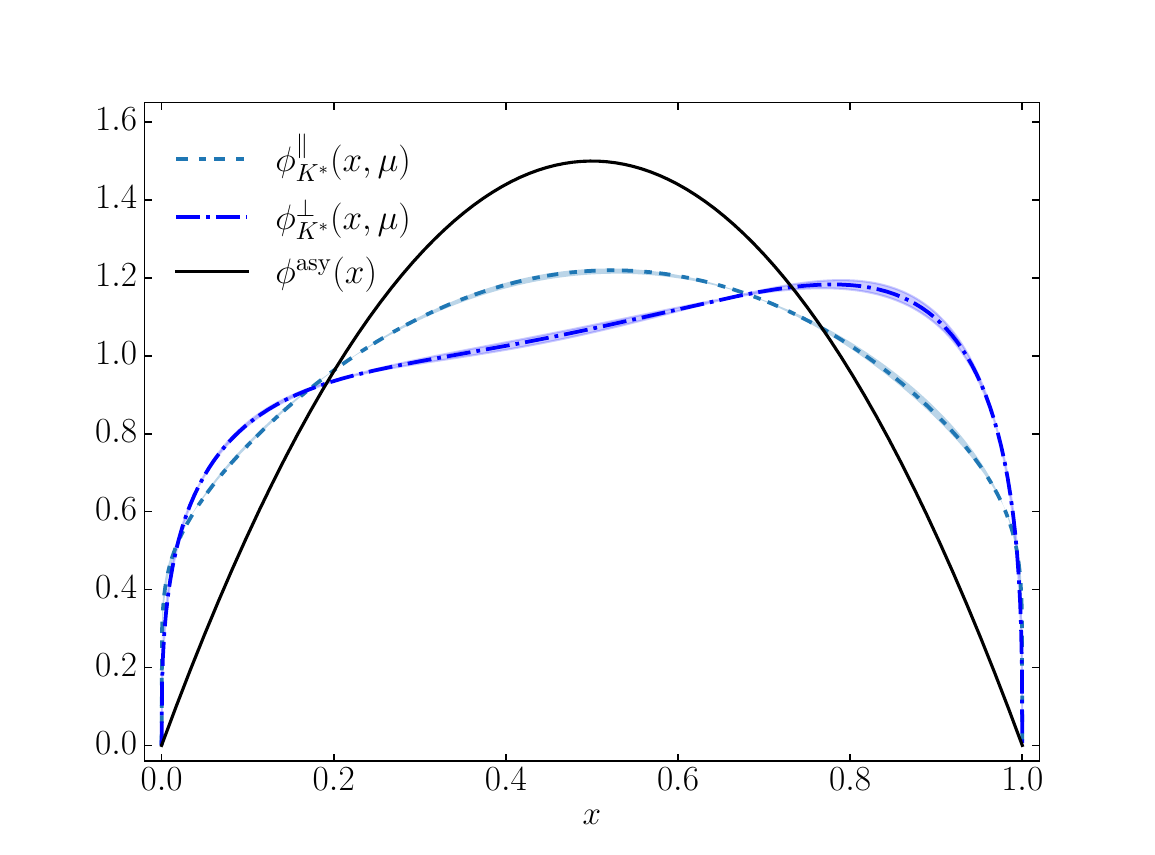}
  \includegraphics[scale=0.434,angle=0]{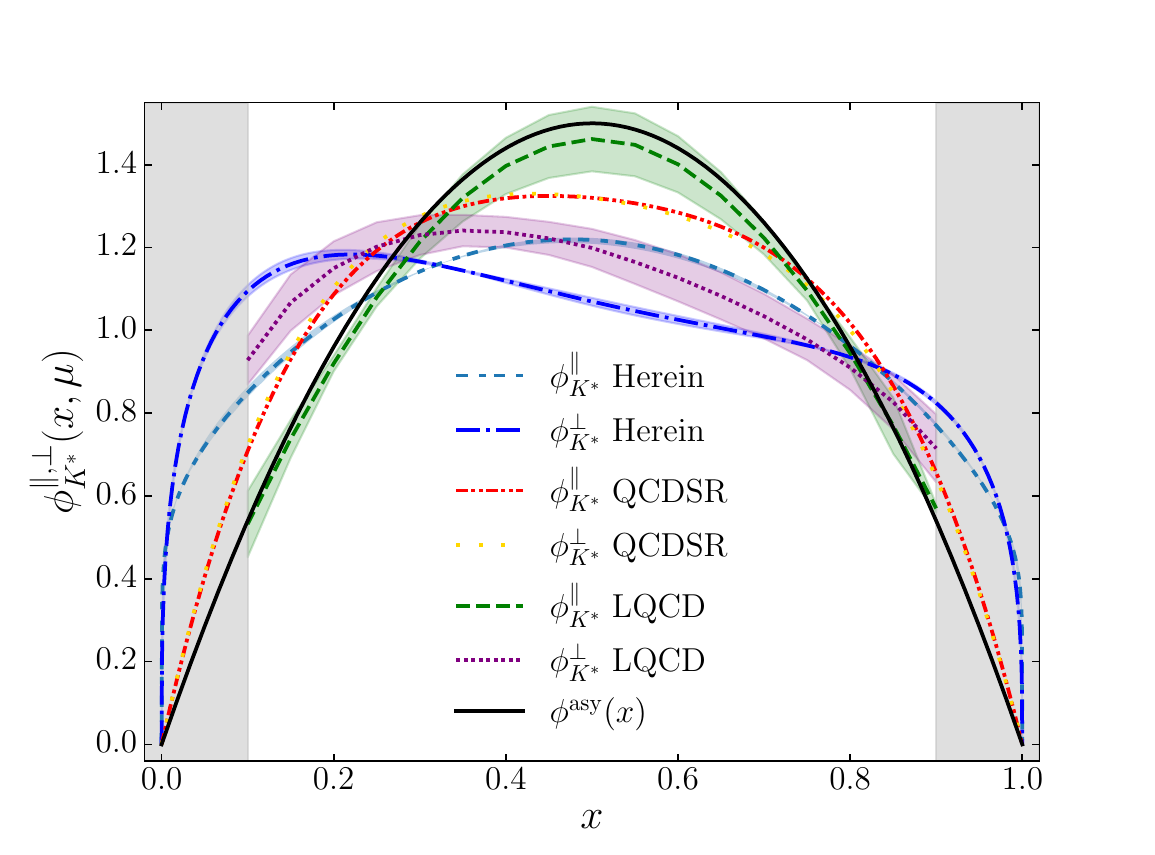} 
 \caption{
 \emph{Top panel\/}:  longitudinal and transverse distribution amplitudes, $\phi^{\parallel}_{K^*}(x,\mu)$ and $\phi^{\perp}_{K^*}(x,\mu)$ for $\mu=2$~GeV. 
 \emph{Bottom panel\/}:  Comparison of our predictions for the $K^*$ with those of QCDSR~\cite{Ball:2007zt} and LQCD~\cite{Hua:2020gnw},
 where we replaced $x\rightarrow1-x$ in Eq.~\eqref{reco}. The shaded areas and error bands are as in Figure~\ref{fig:Compa-LCDAs-rho-phi}.}
 \label{fig:Compa-LCDAs-Kstar} 
\end{figure}
%
\vspace*{-1cm}

\section{Results}

We begin with the light vector mesons and determine the coefficients $a_n^{\parallel,\perp}$ of their Gegenbauer expansion via a least-square fit of 
$\epsilon_{ \parallel,\perp}$~\eqref{reco} with the four moments $\langle  x \rangle,  \langle  x^2 \rangle,  \langle  x^3 \rangle,  \langle  x^4 \rangle$. 
We report their values and those of the corresponding $a_n^{\parallel,\perp}$ of the $\rho$, $\phi$ and $K^*$ mesons in Table~\ref{x-moments-rho-phi-K} 
and compare  the moments $\langle  \xi^{2m} \rangle_{\parallel,\perp}$~\eqref{moments-v2} with other results in Table~\ref{xi-moments-rho-phi-K}.

The LFDAs for the $\rho$ and $\phi$ mesons are compared in Figure~\ref{fig:Compa-LCDAs-rho-phi} with the prediction of a DSE-based calculation and the LDFA
reconstructed with moments from LQCD, respectively. We infer that the distributions follow the expected pattern: both LFDAs are symmetric about the midpoint, 
$x=1/2$.  However, the $\phi_\rho^ {\parallel,\perp}(x,\mu)$ distributions are broad while $\phi_\phi^ {\parallel,\perp}(x,\mu)$ tend to the asymptotic form 
$\phi(x) \stackrel{\mu \to \infty }{=}  6x\bar x$. In addition, we observe that $\phi^{\parallel}_\rho(x,\mu)$ is slightly broader than $\phi^{\perp}_\rho(x,\mu)$, 
the origin of which are the different values of $a^{\parallel}_2$ and $a^{\perp}_2$ in Table~\ref{x-moments-rho-phi-K}. It appears from Table~\ref{xi-moments-rho-phi-K}
that our calculated  $\langle \xi^{2m} \rangle_{\parallel}$ moments for the $\rho$ meson are overall about 11\% larger, whereas the values for $\langle \xi^{2m} \rangle_{\perp}$ 
are in very good agreement with those of Ref.~\cite{Gao:2014bca} and the HERA fit~\cite{Forshaw:2010py,Forshaw:2012im}.

In the case of the $\phi$-meson, we note  that $\phi^{\parallel}_\phi(x,\mu)\approx\phi^{\perp}_\phi(x,\mu)$ since $a^{\parallel}_2\approx a^{\perp}_2$ and 
$\alpha^\parallel \approx \alpha^\perp$. We remark that our results for $\phi_\phi^ {\parallel,\perp}(x,\mu)$ differ from those in Ref.~\cite{Gao:2014bca} as can 
be inferred from Fig.~\ref{fig:Compa-LCDAs-rho-phi}. The reason for this, despite a like-minded BSE approach, is that we use a larger strange-quark mass, 
$m_s = 166$~MeV at $\mu =2$~GeV. With a lower value of $m_s \approx 100$~MeV we find similar distributions as in Ref.~\cite{Gao:2014bca}. However, 
we prefer to renormalize the DSE with a larger strange mass as it results in a more consistent description of the $K$, $K^*$ and $\phi$ mesons.

%
\begin{table*}[t!]
\centering
\renewcommand{\arraystretch}{1.3}
\setlength{\tabcolsep}{10pt}

\begin{tabular}{c|c|c|c|c||c|c}
\hline
   & $\langle x\rangle_{\parallel,\perp}$  &  $\langle x^2\rangle_{\parallel,\perp}$  &  $\langle x^3\rangle_{\parallel,\perp}$  &  $\langle x^4\rangle_{\parallel,\perp}$ &
      $\alpha_{\parallel,\perp}$  &  $\beta_{\parallel,\perp}$  \\ 
\hline
  $J/\Psi_\parallel$ &0.500&0.274& 0.159 & 0.097 & $4.549\pm0.411$ & $0.081\pm 0.051$  \\
  $J/\Psi_\perp$  & 0.500& 0.259& 0.139& 0.076 & $12.703\pm1.931$ & $0.004\pm 0.710$  \\ \hline
  $D^*_\parallel$ &0.694 & 0.511& 0.396 & 0.315 & $0.531\pm 0.207$ & $ 2.460\pm0.131$   \\
  $D^*_\perp$ &  0.742 & 0.589 & 0.471 & 0.389 & $0.094\pm 0.001$ & $3.073\pm0.001$      \\ \hline
  $D^*_{s\parallel}$ & 0.627 & 0.418 & 0.294 & 0.217 & $2.582\pm0.651$ & $2.263\pm0.296$  \\
  $D^*_{s\perp}$ & 0.655 & 0.465 & 0.346 & 0.272 & $0.448\pm0.305$&$1.832\pm0.136$   \\
\hline 
\end{tabular}
\caption{Mellin moments $\langle x^m\rangle_{\parallel}$ and $\langle x^m\rangle_{\perp}$ of the $J/\Psi$, $D^*$ and $D^*_s$ mesons. 
             Fitting these moments with their definitions in Eqs.~\eqref{xmpara} and \eqref{xmperp} and the corresponding $\phi_{V}^{\parallel,\perp}(x,\mu)$  
             parametrization~\eqref{phi-heavy} yields $\alpha_{\parallel,\perp}$ and $\beta_{\parallel,\perp}$; the fit errors arise in the minimization process. }
\label{LCDAs-heavy}
\end{table*}

\begin{figure}[b!] 
\vspace*{-7mm}
\centering
  \includegraphics[scale=0.55,angle=0]{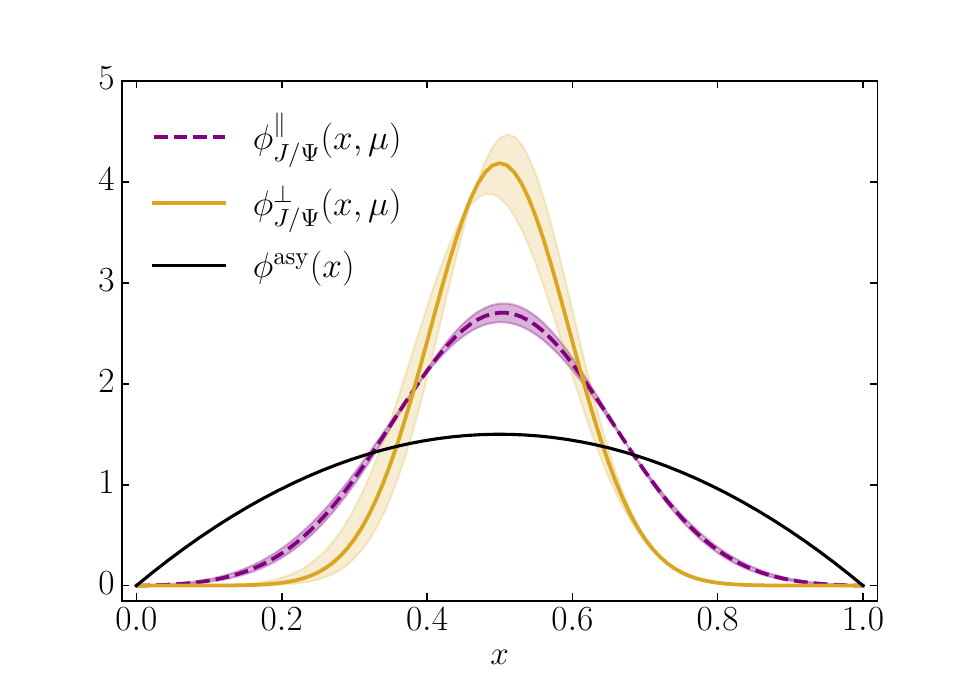}  
 \caption{Longitudinal and transverse distributions, $\phi^{\parallel}_{J/\Psi}(x,\mu)$ and $\phi^{\perp}_{J/\Psi}(x,\mu)$, reconstructed from 
              the moments in Table~\ref{LCDAs-heavy} with Eqs.~\eqref{xmpara}, \eqref{xmperp} and \eqref{phi-heavy}. The error bands reflect the uncertainties 
              in the fit parameters  $\alpha_{\parallel,\perp}$ and  $\beta_{\parallel,\perp}$  in Table~\ref{LCDAs-heavy}.}%
 \label{LCDAs-Jpsi} 
\end{figure}

We now turn our attention to the $K^*$ and present the longitudinal and transverse LFDAs in Figure~\ref{fig:Compa-LCDAs-Kstar} where we juxtapose them with predictions 
from LQCD and QCDSR. Notably, the longitudinal distribution is a concave, nearly symmetric function of $x$, much broader than the asymptotic form, which is a 
consequence of the smallness of the $a^\parallel_1$ coefficient. The transverse LFDA, on the other hand, is asymmetric around the midpoint and its maximum is 
located at $x = 0.78$, which clearly indicates SU(3) flavor symmetry breaking and that the strange valence quark carries a larger amount of meson momentum.  
The asymmetric shape is due to the similarity of the Gegenbauer coefficients, $a^\perp_1 \approx a^\perp_2$ whereas  $a^\parallel_1 \ll a^\parallel_2$, see 
Table~\ref{x-moments-rho-phi-K}.  This is in agreement with a recent calculation in LQCD, though in that study $\phi^\parallel_{K^*}(x,\mu)$ tends toward the asymptotic 
distribution~\cite{Hua:2020gnw}. In contrast to these findings, QCDSR predicts $\phi^\parallel_{K^*}(x,\mu) \approx \phi^\perp_{K^*}(x,\mu)$~\cite{Ball:2007zt}.

As we noted earlier, the heavier vector charmonium and charmed mesons require a modified description of their LFDA~\eqref{phi-heavy} to fit the moments. 
We report these moments, $\langle x^m\rangle_{\parallel}$ and $\langle x^m\rangle_{\perp}$, for the $J/\Psi$, $D^*$ and $D^*_s$ in Table~\ref{LCDAs-heavy}.
The distributions $\phi^{\parallel}_{J/\Psi}(x,\mu)$ and $\phi^{\perp}_{J/\Psi}(x,\mu)$ we then reconstruct are plotted in Figure~\ref{LCDAs-Jpsi}. They are reminiscent 
of their pseudoscalar counterpart, i.e. the LFDA of the $\eta_c$, which exhibits the same \emph{convex-concave-convex\/} functional behavior and is more sharply 
peaked than the asymptotic LFDA~\cite{Serna:2020txe}. It turns out that the longitudinal distribution is broader and less localized as a function of $x$ than the transverse 
distribution, an observation also made in Ref.~\cite{Ding:2015rkn}.

\begin{figure}[b!] 
\vspace*{-7.5mm}
\centering
  \includegraphics[scale=0.52,angle=0]{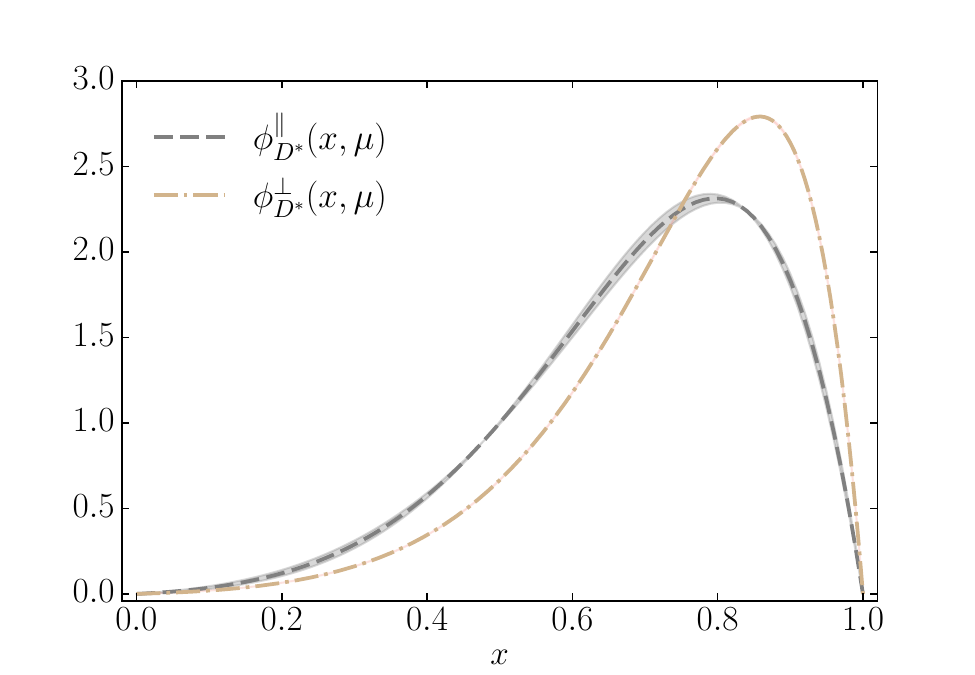}
  \includegraphics[scale=0.52,angle=0]{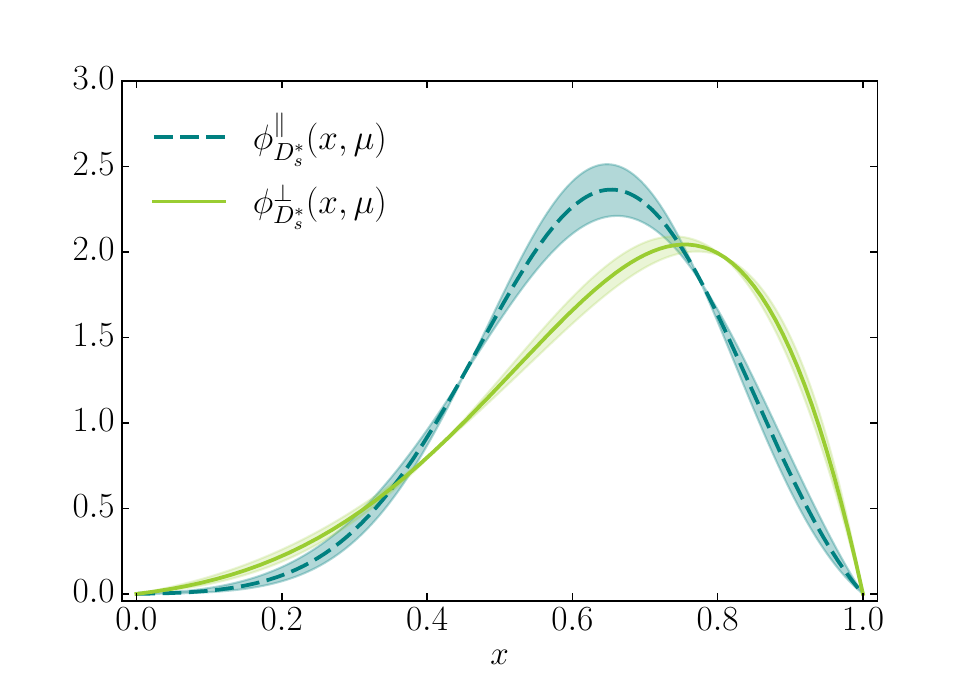} 
 \caption{Longitudinal and transverse LFDAs of the $D^*$ and $D_s^*$ mesons at $\mu=2$~GeV; error bands as in Figure~\ref{LCDAs-Jpsi}.}
\label{LCDAs-Charmed}   
\end{figure}

We conclude this section with a first prediction of the $D^*$ and $D_s^*$ meson distribution amplitudes which we compute with the projections in Eqs.~\eqref{Par-LCDA} 
and  \eqref{Perp-LCDA} of the Bethe-Salpeter wave functions. The latter are taken from Ref.~\cite{El-Bennich:2021ldv}; see Table~4 therein for the corresponding masses 
and weak decay constants. The distributions we reconstruct from $\langle x^m\rangle_{\parallel}$ and $\langle x^m\rangle_{\perp}$ listed in Table~\ref{LCDAs-heavy}
are shown in Fig.~\ref{LCDAs-Charmed}. Clearly, in both cases the LFDAs are asymmetric and $\phi^{\parallel}_{D^*}(x,\mu)$ and $\phi^{\perp}_{D^*}(x,\mu)$ peak 
at about $x\approx 0.8-0.85$, while $\phi^{\parallel}_{D^*_s}(x,\mu)$ and $\phi^{\perp}_{D^*_s}(x,\mu)$ reach their maximum in the range  $x\approx 0.65 - 0.8$.

This is readily interpreted as the charm quark carrying most of the light-front momentum in the $D^*$ meson, but less so in the $D^*_s$ meson.
Interestingly, the transverse distributions are more asymmetric and the charm seems to carry a larger fraction of the meson momentum than in the longitudinal
distribution. Arguably, this observation generalizes our results for the $K^*$, where the much smaller mass difference between the strange and up quarks leads 
to an almost symmetric form of $\phi^{\parallel}_{K^*}(x,\mu)$ and to a broad yet asymmetric function $\phi^{\perp}_{K^*}(x,\mu)$.


\section{Conclusion}

We extracted the LFDAs of the $\rho$, $\phi$, $K^*$, $J/\Psi$, $D^*$ and $D^*_s$ mesons from their Bethe-Salpeter wave functions, which we calculated in 
Refs.~\cite{Mojica:2017tvh,El-Bennich:2021ldv}, with two projections onto the light front given by Eqs.~\eqref{Par-LCDA} and~\eqref{Perp-LCDA}. The transverse LFDA 
of the $\rho$ meson is in very good agreement with that obtained in a similar DSE-BSE approach~\cite{Gao:2014bca} and with the HERA fit~\cite{Forshaw:2010py,
Forshaw:2012im}, while our longitudinal moments, $\langle \xi^m \rangle_\parallel$, are generally about 11\% larger than those in the literature. 

We then presented the first calculation of the $\phi^{\parallel}_{K^*}(x,\mu)$ and $\phi^{\perp}_{K^*}(x,\mu)$ within the DSE-BSE framework and confirm the functional 
form  found with LQCD simulations~\cite{Hua:2020gnw}: while the longitudinal distribution of the $K^*$ is almost symmetric about 
the midpoint $x=1/2$, the transverse distribution is broad and slanted, which we interpret as the strange quark carrying the larger fraction of the meson's momentum. 
In the heavy meson sector, both LFDAs of the $J/\Psi$ are alike with that of the $\eta_c$, i.e. they are symmetric and narrow, yet not merely concave distributions.

Last not least, we extended our studies in Ref.~\cite{Serna:2020txe} to the longitudinal and transverse LFDAs of the $D^*$ and $D_s^*$ mesons, a first calculation
of these distributions to our knowledge. Our findings are in line with observations for the pseudoscalar $D$ and $D_s$ mesons~\cite{Serna:2020txe}: the distributions 
are asymmetric and reach their maximum at large momentum fractions, namely $x\approx 0.65 - 0.85$. In other words, the charm quark is most likely to carry the largest 
fraction of  the $D_{(s)}^*$-momentum, and this is even more so the case for the transverse distribution. 

We remind that we provided all the analytic parametrizations of the LFDAs discussed in this work and the parameters are found in Tables~\eqref{x-moments-rho-phi-K}       
and \eqref{LCDAs-heavy}. Therefore, the LFDAs of the $J/\Psi$ and $D^*_{(s)}$ mesons can readily be used in diffractive vector-meson production and are are of interest
to the experimental program of the Electron-Ion Collider.


\section*{Acknowledgments} 

We acknowledge helpful discussions with Peter Tandy and Minghui Ding.
B.E. and F.E.S. participate in the Brazilian network project \emph{INCT-F\'isica Nuclear e Aplica\c{c}\~oes\/}, no.~464898/2014-5. This work was supported by 
the S\~ao Paulo Research Foundation (FAPESP), grant no.~2018/20218-4, and by the National Council for Scientific and Technological Development (CNPq), 
grant no.~428003/2018-4.  F.E.S. is a CAPES-PNPD postdoctoral fellow financed by grant no.~88882.314890/2013-01.



\begin{thebibliography}{99}

\bibitem{dePaula:2020qna}
W.~de Paula, E.~Ydrefors, J.~H.~Alvarenga Nogueira, T.~Frederico and G.~Salm\`e,
Phys. Rev. D \textbf{103} (2021) no.1, 014002
doi:10.1103/PhysRevD.103.014002

\bibitem{Efremov:1978rn}
A.~V.~Efremov and A.~V.~Radyushkin,
Theor. Math. Phys. \textbf{42} (1980), 97-110
doi:10.1007/BF01032111

\bibitem{Lepage:1980fj}
G.~P.~Lepage and S.~J.~Brodsky,
Phys. Rev. D \textbf{22} (1980), 2157
doi:10.1103/PhysRevD.22.2157

\bibitem{Chang:2013nia}
L.~Chang, I.~C.~Clo\"et, C.~D.~Roberts, S.~M.~Schmidt and P.~C.~Tandy,
Phys. Rev. Lett. \textbf{111} (2013) no.14, 141802
doi:10.1103/PhysRevLett.111.141802

\bibitem{Raya:2015gva}
K.~Raya, L.~Chang, A.~Bashir, J.~J.~Cobos-Mart\'inez, L.~X.~Guti\'errez-Guerrero, C.~D.~Roberts and P.~C.~Tandy,
Phys. Rev. D \textbf{93} (2016) no.7, 074017
doi:10.1103/PhysRevD.93.074017

\bibitem{Raya:2019dnh}
K.~Raya, A.~Bashir and P.~Roig,
Phys. Rev. D \textbf{101} (2020) no.7, 074021
doi:10.1103/PhysRevD.101.074021

\bibitem{Beneke:2002jn}
M.~Beneke and M.~Neubert,
Nucl. Phys. B \textbf{651} (2003), 225-248
doi:10.1016/S0550-3213(02)01091-X

\bibitem{Bauer:2005wb}
C.~W.~Bauer, D.~Pirjol, I.~Z.~Rothstein and I.~W.~Stewart,
Phys. Rev. D \textbf{72} (2005), 098502
doi:10.1103/PhysRevD.72.098502

\bibitem{El-Bennich:2006rcn}
B.~El-Bennich, A.~Furman, R.~Kaminski, L.~Lesniak and B.~Loiseau,
Phys. Rev. D \textbf{74} (2006), 114009
doi:10.1103/PhysRevD.74.114009

\bibitem{El-Bennich:2009gqk}
B.~El-Bennich, A.~Furman, R.~Kaminski, L.~Lesniak, B.~Loiseau and B.~Moussallam,
Phys. Rev. D \textbf{79} (2009), 094005
[erratum: Phys. Rev. D \textbf{83} (2011), 039903]
doi:10.1103/PhysRevD.83.039903

\bibitem{Leitner:2010fq}
O.~Leitner, J.~P.~Dedonder, B.~Loiseau and B.~El-Bennich,
Phys. Rev. D \textbf{82} (2010), 076006
doi:10.1103/PhysRevD.82.076006

\bibitem{Pire:2015iza}
B.~Pire and L.~Szymanowski,
Phys. Rev. Lett. \textbf{115} (2015) no.9, 092001
doi:10.1103/PhysRevLett.115.092001

\bibitem{Pire:2017yge}
B.~Pire and L.~Szymanowski,
Phys. Rev. D \textbf{96} (2017) no.11, 114008
doi:10.1103/PhysRevD.96.114008

\bibitem{Pire:2017lfj}
B.~Pire, L.~Szymanowski and J.~Wagner,
Phys. Rev. D \textbf{95} (2017) no.9, 094001
doi:10.1103/PhysRevD.95.094001

\bibitem{Pire:2021dad}
B.~Pire, L.~Szymanowski and J.~Wagner,
Phys.  Rev. D \textbf{104} (2021) no.9, 094002
doi:10.1103/PhysRevD.104.094002

\bibitem{Chang:2013pq}
L.~Chang, I.~C.~Clo\"et, J.~J.~Cobos-Mart\'inez, C.~D.~Roberts, S.~M.~Schmidt and P.~C.~Tandy,
Phys. Rev. Lett. \textbf{110} (2013) no.13, 132001
doi:10.1103/PhysRevLett.110.132001

\bibitem{Shi:2015esa}
C.~Shi, C.~Chen, L.~Chang, C.~D.~Roberts, S.~M.~Schmidt and H.~S.~Zong,
Phys. Rev. D \textbf{92} (2015), 014035
doi:10.1103/PhysRevD.92.014035

\bibitem{Serna:2020txe}
F.~E.~Serna, R.~C.~da Silveira, J.~J.~Cobos-Mart\'\i{}nez, B.~El-Bennich and E.~Rojas,
Eur. Phys. J. C \textbf{80} (2020) no.10, 955
doi:10.1140/epjc/s10052-020-08517-3

\bibitem{Beneke:2000wa}
M.~Beneke and T.~Feldmann,
Nucl. Phys. B \textbf{592} (2001), 3-34
doi:10.1016/S0550-3213(00)00585-X

\bibitem{Forshaw:2010py}
J.~R.~Forshaw and R.~Sandapen,
JHEP \textbf{11} (2010), 037
doi:10.1007/JHEP11(2010)037

\bibitem{Forshaw:2012im}
J.~R.~Forshaw and R.~Sandapen,
Phys. Rev. Lett. \textbf{109} (2012), 081601
doi:10.1103/PhysRevLett.109.081601

\bibitem{Bashir:2012fs}
A.~Bashir, L.~Chang, I.~C.~Cloet, B.~El-Bennich, Y.~X.~Liu, C.~D.~Roberts and P.~C.~Tandy,
Commun. Theor. Phys. \textbf{58} (2012), 79-134
doi:10.1088/0253-6102/58/1/16

\bibitem{Gao:2014bca}
F.~Gao, L.~Chang, Y.~X.~Liu, C.~D.~Roberts and S.~M.~Schmidt,
Phys. Rev. D \textbf{90} (2014) no.1, 014011
doi:10.1103/PhysRevD.90.014011

\bibitem{Ding:2015rkn}
M.~Ding, F.~Gao, L.~Chang, Y.~X.~Liu and C.~D.~Roberts,
Phys. Lett. B \textbf{753} (2016), 330-335
doi:10.1016/j.physletb.2015.11.075

\bibitem{Ball:2007zt}
P.~Ball, V.~M.~Braun and A.~Lenz,
JHEP \textbf{08} (2007), 090
doi:10.1088/1126-6708/2007/08/090

\bibitem{Boyle:2008nj}
P.~A.~Boyle \textit{et al.} [RBC and UKQCD],
PoS \textbf{LATTICE2008} (2008), 165
doi:10.22323/1.066.0165

\bibitem{Hua:2020gnw}
J.~Hua \textit{et al.} [Lattice Parton],
Phys. Rev. Lett. \textbf{127} (2021) no.6, 062002
doi:10.1103/PhysRevLett.127.062002

\bibitem{Lu:2021sgg}
Y.~Lu, D.~Binosi, M.~Ding, C.~D.~Roberts, H.~Y.~Xing and C.~Xu,
Eur. Phys. J. A \textbf{57} (2021) no.4, 115
doi:10.1140/epja/s10050-021-00427-6

\bibitem{Mojica:2017tvh}
F.~F.~Mojica, C.~E.~Vera, E.~Rojas and B.~El-Bennich,
Phys. Rev. D \textbf{96} (2017) no.1, 014012
doi:10.1103/PhysRevD.96.014012

\bibitem{El-Bennich:2021ldv}
B.~El-Bennich and F.~E.~Serna,
PoS \textbf{CHARM2020} (2021), 025
doi:10.22323/1.385.0025

\bibitem{Gao:2016jka}
F.~Gao, L.~Chang and Y.~x.~Liu,
Phys. Lett. B \textbf{770} (2017), 551-555
doi:10.1016/j.physletb.2017.04.077

\bibitem{Serna:2021xnr}
F.~E.~Serna and B.~El-Bennich,
PoS \textbf{CHARM2020} (2021), 047
doi:10.22323/1.385.0047

\end{thebibliography}
\end{document}